\documentclass[apj]{emulateapj}

\usepackage{graphicx}
\usepackage{txfonts}
\usepackage{lscape}
\usepackage{aas_macros}
\usepackage{amssymb}
\usepackage{natbib}
\bibpunct{(}{)}{;}{a}{,}{,}
\usepackage{longtable}
\usepackage[hang,bf,footnotesize]{subfigure}
\usepackage[figuresright]{rotating}

\begin{document}

\title{Optical photometric GTC/OSIRIS observations of the young massive association Cygnus~OB2}

\author{M. G. Guarcello\altaffilmark{1}, N. J. Wright\altaffilmark{1}, J. J. Drake\altaffilmark{1}, D. Garc\'{i}a-Alvarez\altaffilmark{2,3,4}, J. E. Drew\altaffilmark{5}, T. Aldcroft\altaffilmark{1}, V. L. Kashyap\altaffilmark{1}}

\altaffiltext{1}{Smithsonian Astrophysical Observatory, MS-67, 60 Garden Street, Cambridge, MA 02138, USA}
\altaffiltext{2}{Instituto de Astrof\'{i}sica de Canarias, E-38205 La Laguna, Tenerife, Spain}
\altaffiltext{3}{Dpto. de Astrof\'{\i}sica, Universidad de La Laguna, 38206 - La Laguna, Tenerife, Spain}
\altaffiltext{4}{Grantecan CALP, 38712 Bre{\~na} Baja, La Palma, Spain}
\altaffiltext{5}{CAR/STRI, University of Hertfordshire, College Lane, Hatfield, AL10 9AB, UK}

\begin{abstract}
In order to fully understand the gravitational collapse of molecular clouds, the star formation process and the evolution of circumstellar disks, these phenomena must be studied in different Galactic environments with a range of stellar contents and positions in the Galaxy. The young massive association Cygnus~OB2, in the Cygnus-X region, is an unique target to study how star formation and the evolution of circumstellar disks proceed in the presence of a large number of massive stars. We present a catalog obtained with recent optical observations in $r,\,i,\,z$ filters with OSIRIS, mounted on the $10.4\,m$ GTC telescope, which is the deepest optical catalog of Cyg~OB2 to date. \par
	The catalog consist of 64157 sources down to $M=0.15\,M_{\odot}$ at the adopted distance and age of Cyg~OB2. A total of 38300 sources have good photometry in all three bands. We combined the optical catalog with existing X-ray data of this region, in order to define the cluster locus in the optical diagrams. The cluster locus in the $r-i$ vs. $i-z$ diagram is compatible with an extinction of the optically selected cluster members in the $2.64^m<A_V<5.57^m$ range. We derive an extinction map of the region, finding a median value of $A_V=4.33^m$ in the center of the association, decreasing toward the north-west. In the color-magnitude diagrams, the shape of the distribution of main sequence stars is compatible with the presence of an obscuring cloud in the foreground at $\sim850\pm 25\,pc$ from the Sun.

\end{abstract}



\section{Introduction}

	In recent years our knowledge of the star formation process and the early phase of stellar evolution has increased remarkably.  This progress has been driven by a large number of deep spectroscopic and photometric observations at high spatial resolution of star forming regions in our Galaxy and in nearby galaxies.  High performance telescopes, such as the NASA Hubble and Spitzer Space Telescopes, and more recently the ESA Herschel Space Observatory, allow us to study the formation of low- and high-mass stars, the evolution of circumstellar disks, and the collapse of protostellar cores in unprecedented detail. X-ray observations with space telescopes such as {\it Chandra} and XMM-Newton have also been crucial, unveiling the populations of star forming regions down to sub-solar masses, and probing the high energy processes in young stars. \par
Despite this progress, few single star forming regions in our Galaxy have provided the opportunity to study star and planet formation over a large range of stellar masses and in presence of very massive stars, which affect the star formation process and the evolution of their parental cloud mostly thank to their intense ionizing flux. The study of such massive young clusters, in fact, is hampered by their large distance from the Sun, with only a few moderately massive clusters closer than $1.5\,kpc$.  \par
	One exception that has only fairly recently been recognized is the Cygnus~OB2 association \citep{Red67}, in the Cygnus-X giant molecular cloud.  It is the richest and most massive OB2 association within $2\,kpc$ of the Sun.
It harbors a large number of massive stars and an extensive young population of low-mass pre-main sequence stars, likely formed in different episodes. The first spectroscopic study devoted to the massive members of this region \citep{Red67} identified about 300 OB members, and in subsequent studies this estimate has increased. In a 2MASS study, \citet{Kno00} used statistical arguments to suggest the presence of 2600 OB stars and 120 O stars and to estimate a total mass of $\left(4-10 \right) \times 10^4\, M_{\odot}$.  This mass led the author to the conclusion that Cyg~OB2 is a young globular cluster in the Milky Way. A near-infrared spectroscopic survey by \citet{Come02} found slightly fewer O stars but confirmed this conclusion.  While \citet[][see also \citealt{Dre08}]{Han03} finds a somewhat lower total mass. The enormous scale of Cyg~OB2 is exemplified by it containing two of the few O3 stars known in our Galaxy \citep{Wal73}, together with B supergiant stars \citep{Mas91,Negu08}.  Other massive objects, such as Wolf-Rayet stars \citep{Nie98} and the Luminous Blue Variable G79.29+0.46 \citep{Hig94} are found in the Cygnus-X complex in the proximity of the association. \par
	The first studies of the intermediate- and low-mass population of the association counted several thousands of candidate members \citep{Red67,Kno00}, but most recent and reliable estimates based on {\it Chandra} X-ray observations of the central cluster indicate a population of 1000-1500 low-mass members down to subsolar masses \citep{Alba07,Wri09}, among which there are several strong $H\alpha$ emitting objects \citep{Vink08}. \par
The first estimate of the distance to Cyg~OB2 was $2.1\,kpc$ \citep{Red67}. \citet{Mas91} found $1.8\,kpc$ based on a combined photometric and spectroscopic study, while the most recent estimate based on spectroscopy of the most massive members of Cyg~OB2 is $1.45\,kpc$ \citep{Han03}. \citet{Rygl11} reported a distance of the Cygnus-X complex of $1400\,pc$, derived from the trigonometric parallaxes of five star forming region in the complex. We adopt a distance of $1450\,pc$ in this work.  Cyg~OB2 then lies behind the nebulosity associated with the Great Cygnus Rift and, despite its relative proximity to the Sun, the association is significantly absorbed. \par 
	A picture has emerged in the last few years of different episodes of star formation in and around the main center of Cyg~OB2. \citet{Han03} noted the presence of both massive stars younger than $\sim2\,Myr$, and of B stars that appeared older for the adopted association distance, or would need to be in the foreground. \citet{Dre08} identified a population of A-type stars south of the main association with an age of $5-7\,Myr$. Since then, \citet{Wri00} have shown that the NIR color-magnitude diagram of the region indicates both a $\sim3\,Myr$ and a $\sim5\,Myr$ old population of stars down to stellar masses of $\sim1\,M_{\odot}$, while \citet{Wright12} discovered a population of candidate proplyds\footnote{Stars with protoplanetary disks surrounded by a photoevaporating envelope of gas \citep{Dell94}} southward Cyg~OB2 suggesting that new episodes of star formation occurred. \par
In this paper we present the deepest and highest spatial resolution optical observations of Cyg~OB2 to date. In sections \ref{inst_sect} and \ref{phot_sect} we describe the instrument setup, the observations, the data reduction, and the photometric calibration in the AB system; in section \ref{cat_sect} we describe the final catalog, consisting of 64157 optical sources, among which 38300 have good photometry in all three bands. In section \ref{match_sec} we cross-correlate the OSIRIS catalog with the IPHAS data and with the newly released SDSS DR8 data of this region, and with the existing {\it Chandra}/ACIS-I catalog of Cyg~OB2. Finally, in sections \ref{av_sec} and \ref{cm_sec} we present the color-color and color-magnitude diagrams and we use them to study the extinction and age of association members and the distance of the foreground nebulosity. In a forthcoming paper, the optical catalog described in this paper will be combined with new deep X-ray observations of Cyg~OB2 (the $1.08\,Msec$ {\it Chandra} Cygnus~OB2 Legacy Survey, P.I. J.~Drake\footnote{http://www.cygob2.org/}), to obtain the deepest list of stars associated with this unique star forming region. \par

\section{Instrument setup and observations}
\label{inst_sect}

The observations were performed in the $r^{\prime}$, $i^{\prime}$, and $z^{\prime}$ filters with the Optical System for Imaging and low Resolution Integrated Spectroscopy (OSIRIS), mounted on the $10.4\,m$ Gran Telescopio CANARIAS (GTC) of the Spanish Observatorio del Roque de los Muchachos in La Palma \citep{Cepa00}. The OSIRIS detector comprises two Marconi CCDs, each with $2048\times 4096$ pixels, separated along the long side by a 72 pixel gap. The total Field of View (FoV) is $7.8^{\prime} \times 8.5^{\prime}$ ($7.8^{\prime} \times 7.8^{\prime}$ unvignetted), corresponding to a spatial scale of $0.127^{\prime \prime}/pix$. Such a small spatial scale is necessary to resolve very crowded regions such as Cygnus OB2. \par
	OSIRIS provides different binning and frame readout speeds, each corresponding to different values of readout noise (RON) and detector gain\footnote{on-line manual at http://www.gtc.iac.es/en/pages/instrumentation/osiris.php.}. Our observations have been performed with the $1\times 1$ binning mode and with different readout speeds: 100kHz/HN; 500kHz/HN and 500kHz/LN\footnote{HN: High Noise, LN: Low Noise}. Table \ref{log_tbl} summarizes the seeing conditions, configurations and observed Cyg~OB2 and standard fields for each night of observation. The seeing conditions were very good for the first, fourth and fifth nights; acceptable for the other nights. \par
\begin{figure}[!ht]
	\centering	
	\includegraphics[width=8.5cm]{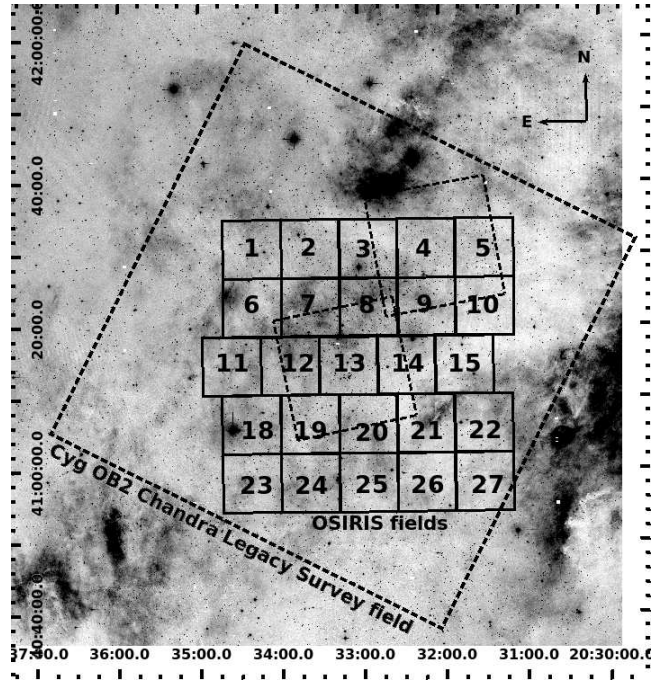}
	\caption{IPHAS $H\alpha$ image of a $1.47^{\circ} \times 1.47^{\circ}$ area centered on Cyg~OB2. The numbered black boxes are the fields observed with OSIRIS. The large inclined dashed box represents the area observed in the {\it Chandra} Cygnus~OB2 Legacy Survey. The small inclined boxes are earlier {\it Chandra} pointings in this region.}
	\label{fov_image}
	\end{figure}

Fig. \ref{fov_image} shows a $1.47^{\circ}\times 1.47^{\circ}$ IPHAS $H\alpha$ image of Cyg~OB2. The OSIRIS observations are arranged in a $5\times 5$ mosaic in order to cover the $41^{\prime} \times 41^{\prime}$ central part of the area imaged in X-rays for the {\it Chandra} Cygnus~OB2 Legacy Survey (large box). Each OSIRIS field is labeled with a number ranging from 1 to 27. In the rest of this paper, we will refer to the fields using these numbers. Fields 16 and 17 are not shown in Fig.~\ref{fov_image} since they were observed in very bad seeing conditions on 8/19/2009 and they have not been analyzed. Each field was observed three times per filter, for a total of 225 images, with small offsets in RA to cover both the gap between the CCDs and vignetted area. In this way, we obtained full coverage of the field in the RA direction, with some gap in the DEC direction due to CCD vignetting. The total exposure time in each field and in each filter was 80 sec, which allowed us to observe stars down to $r^{\prime}\sim25^m$. The small inclined boxes in Fig.~\ref{fov_image} show the area previously observed with {\it Chandra} \citep{Butt06,Alba07}. \par

\section{Data reduction and photometry}
\label{phot_sect}

Overscan, bias, and flat field corrections were made using the IRAF $CCDPROC$ task of the $MSCRED$ package. No correction for dark current was been made since it does not significantly affect the OSIRIS detectors. \par
	Source detection and Point Spread Function (PSF) photometry were performed using the DAOPHOT/ALLSTAR packages \citep{Ste87,Ste94}. For source detection we used a threshold ranging from 3.5 to 4.5 times the standard deviation of the average sky emission level ($\sigma_{sky}$), and the stellar profiles were fitted with a Gaussian function with a Full Width at Half Maximum (FWHM) ranging from $0.9^{\prime \prime}$ to $1.5^{\prime \prime}$, depending on the seeing conditions. The PSF photometry has been performed fitting the observed stellar profiles with different PSF models, described by either a Gaussian, Moffat, Lorentz, or Penny functions.  We also accounted for a possible linear variation of the PSF model with the plate coordinates. \par
In order to combine the different observations of each star, we calculated the plate transformations between the different images of the same field, by using DAOMASTER/DAOMATCH procedures \citep{Ste93}, adopting a linear translation of the coordinate systems. Self-consistent sets of PSF stellar magnitudes and positions between all the images of the same field have been obtained using ALLFRAME \citep{Ste94} and the plate transformations calculated with DAOMASTER/DAOMATCH.  We calculated the aperture correction to the photometry using the DAOGROW routine \citep{Ste90}, which uses the $growth-curves$ method, and a detailed model for the stellar profile. \par

\subsection{Photometry and calibration}
\label{cal_sect}
	
	Photometric calibration in the SDSS {\emph riz} photometric system was carried out using the CCDSTD/CCDAVE/TRIAL procedures \citep{Ste05}. These three routines calculate the coefficients of the chosen photometric solution, then apply the transformation first to the standard stars together with a selected sample of target stars, and then to the whole catalog. We calibrated the catalog in the SDSS $riz$ photometric system since the calibrated magnitudes of the observed standard stars were taken from the eighth data release of the Sloan Digital Sky Survey (SDSS DR8, \citealp{Aihara11}). \par
Table \ref{log_tbl} summarizes the observations of the standard fields. Unfortunately, SDSS DR8 data are not available for the field GD71, meaning that the standard stars necessary for the calibration of the images taken on August 18th were not available. \par
We calibrated the observations of Cyg~OB2 taken on the other nights (all the fields with the exception of 6 and 7) using the SDSS photometry of the standard stars in the observed standard fields. We used a standard photometric solution:

\begin{equation}
O_k = M_k + ZP_k + A1_k \times CT + A2_k \times Q
\label{calib_eq}
\end{equation}

Here, $k$ is one of the photometric bands ($r$, $i$, or $z$), $O_k$ is the instrumental magnitude in the $k$ band, $M_k$ is the calibrated magnitude, $ZP_k$ is the photometric zero-point, $CT$ is a first order color term, and Q is the airmass.  Since the standard stars have not been observed over a wide range of airmass, we set $A2$ equal to the extinction in $r^{\prime}$ measured during the observing nights by the Carlsberg Meridian Telescope\footnote{http://www.ast.cam.ac.uk/ioa/research/cmt/camc\_extinction.html}, converted in $i^{\prime}$ and $z^{\prime}$ using the extinction curve derived by King 1985\footnote{http://www.ing.iac.es/Astronomy/observing/manuals/ps/tech\_notes/tn031.pdf}. We calculated $A1$ from the stars observed in both CCDs, and then two different zero-points from the CCDs separately. Table \ref{coeff_table} shows how the coefficients used for the photometric calibration vary in the 5 observing nights. Residual zero-point and color corrections have been applied for each field after matching the GTC and SDSS DR8 (Sect. \ref{sdss_sec}). \par
	Fields 6 and 7 have been calibrated by taking advantage of the fact that field 8 has been observed both on August 18th and 20th. We simply selected a sample of good stars compatible with the MS falling in this field, and then we used their calibrated photometry to propagate the photometric solution in fields 6 and 7. \par
The final photometric solutions reproduce properly the calibrated magnitudes of standard stars from the instrumental magnitudes, for the whole range of stellar magnitudes, colors, positions and airmass. In all cases, the distributions of the residuals (the difference between the calibrated and standard magnitudes for the standard stars) have medians close to zero and RMS deviations of about, or smaller than, 0.035. \par
	The astrometric solution was found using the IRAF tools $CCXYMATCH$, $SETWCS$ and $SKYPIX$, adopting the 2 Micron All Sky Survey (2MASS) Point Source Catalog (PSC, \citealp{Cutri03}) as astrometric reference and with a $tanx$ projection of the plate coordinates onto the celestial system. \par

\section{The final catalog}
\label{cat_sect}

Table \ref{cat_tab} shows part of the final OSIRIS catalog of Cyg~OB2. For each star it provides an identification number (column 1), the celestial coordinates (columns 2 and 3), magnitudes and errors in $r$, $i$, and $z$ (columns 4 to 9), and the $\chi^2$ and {\tt SHARP} parameters. The former is the chi-square derived in the PSF fitting phase, the latter is a parameter describing the shape of the stellar profile \citep{Ste87}. These parameters lose their full statistical meaning when the PSF models from different images of the same star are combined in the ALLFRAME phase, but they are still useful to select stars with good and bad photometry. \par
	The merged catalog consists of 64157 sources. As explained in Table \ref{cat_tab}, stars with $\chi^2$ and {\tt SHARP} parameters equal to 0 have been detected only in one band ($z^{\prime}$ usually). The magnitudes of these stars are not calibrated properly (they do not have an available color). Since they might be useful for future multi-wavelength studies involving the OSIRIS catalog they are retained in order to preserve the locations of known faint optical sources.  These stars can be easily selected by requiring $\chi^2=0$ (there are 50 of these {\it $null-\chi^2$ stars}). \par
\begin{figure}[]
	\centering	
	\includegraphics[width=8.5cm]{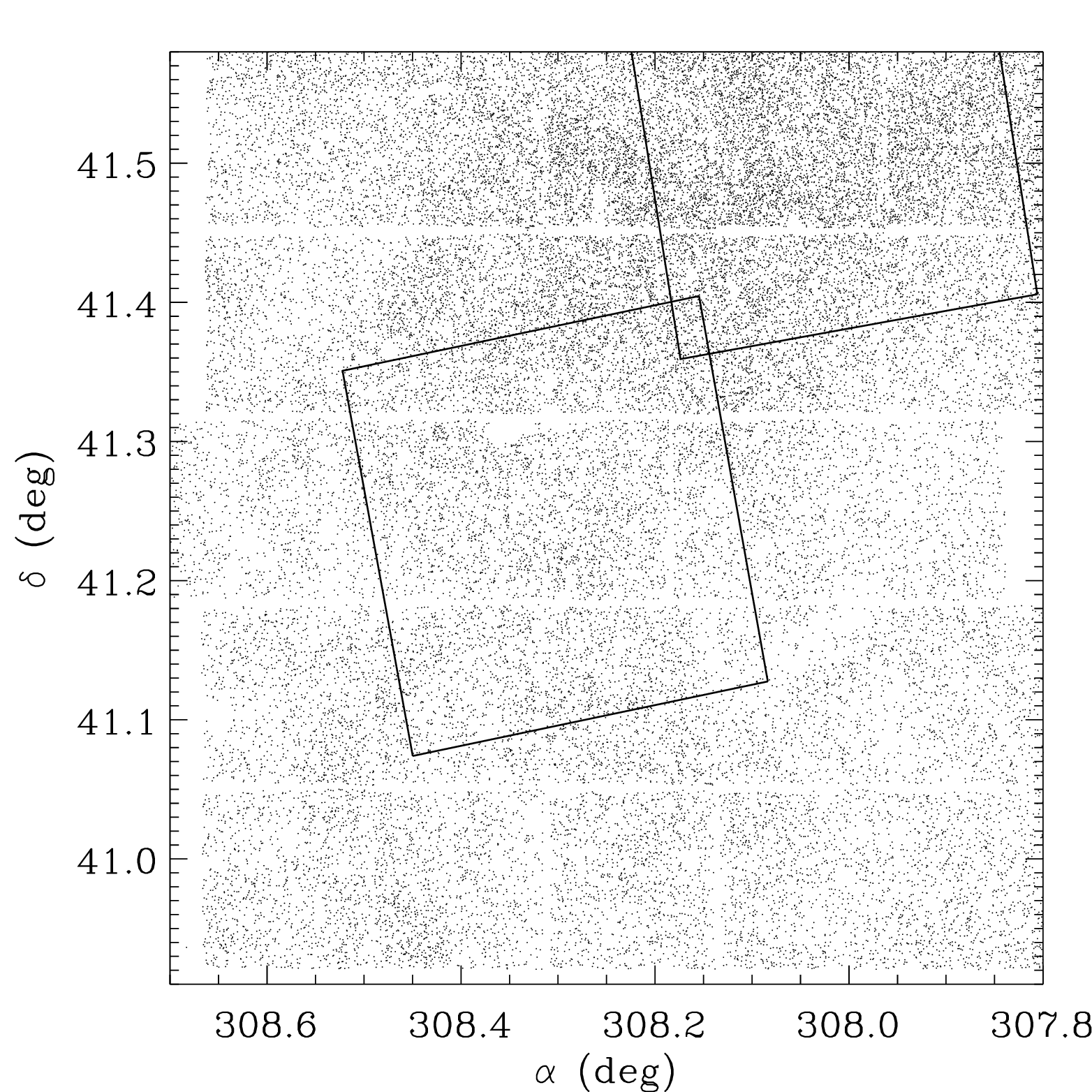}
	\caption{Spatial distributions of the $good-photometry$ sources in the OSIRIS catalog of Cyg~OB2. The boxes delimit the {\it Chandra} ACIS-I pointings presented in \citet{Alba07} and \citet{Butt06}.}
	\label{spadis_image}
	\end{figure}

Stars with bad PSF fitting or with non-circular profiles can be easily selected by requiring $\chi^2>5$ or $|sharp|>8$.  A total of 1815 stars (2.8\% of the whole catalog) satisfy these criteria.  Most of them are sources close to the CCD edges and corners, diffraction spikes created by bright stars, or saturated sources with irregular profiles. The total number of sources with both good shape and $\chi^2$ but with errors in colors larger than 0.15$^m$ is 23992, or 37.4\% of the whole catalog.  Finally, the sources with good photometry in all three bands number 38300, or 59.7\% of the catalog.  We will call these sources $good-photometry$ $sources$ hereafter. The saturation limits have been estimated, by inspection of the stellar profiles, to be $r^{\prime}=16^m$, $i^{\prime}=15.5^m$, and $z^{\prime}=15^m$.
	Fig.~\ref{spadis_image} shows the spatial distribution of the $good-photometry$ stars of the OSIRIS catalog. The gaps in declination are due to the fact that stars falling in vignetted areas do not have the properties required for the $good-photometry$ credential. The two {\it Chandra} ACIS-I pointings of Cyg~OB2 presented by \citet{Alba07} and \citet{Butt06} are also shown. In Fig.~\ref{spadis_image} there is an evident increase in the density of the $good-photometry$ stars from the south to the north, indicating a variation of the visual extinction in the south-north direction. In fact, the comparison of Fig.~\ref{spadis_image} with infrared images of Cyg~OB2 (not shown here) reveals that the area to the north-west, which is that with the highest visible stellar density, corresponds to a cavity in the molecular cloud where a large number of background sources can be observed \citep{Red67}. \par
\begin{figure}[]
	\centering	
	\includegraphics[width=8.5cm]{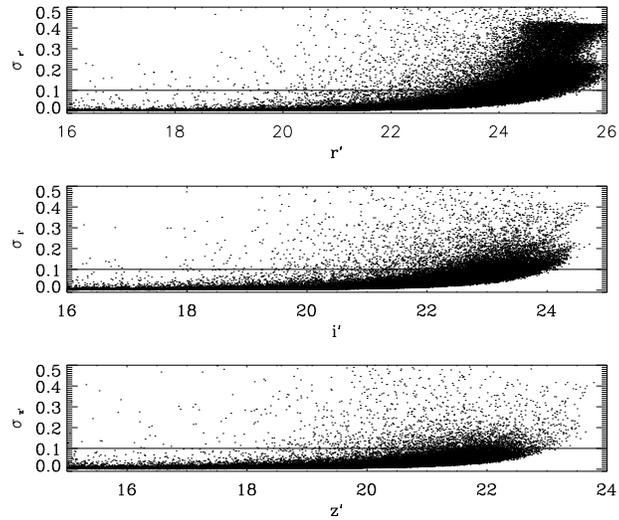}
	\caption{Photometric errors vs. magnitudes in $r^{\prime}$, $i^{\prime}$, and $z^{\prime}$ for all the stars in the OSIRIS catalog. The horizontal line marks the $\sigma=0.1^m$ limit. }
	\label{ermag_image} 
	\end{figure}

Fig.~\ref{ermag_image} shows the photometric errors vs. the magnitude in the three bands. In each panel, the horizontal $\sigma=0.1^m$ line shows the depth of the photometry of the $good-photometry$ stars, reaching $r^{\prime}\sim25^m$, $i^{\prime}\sim24^m$, and  $z^{\prime}\sim22.5^m$.

\section{Catalog cross-matching}
\label{match_sec}

	\subsection{Cross-correlation with the {\it SDSS DR8} catalog}
	\label{sdss_sec}

The data from the Eighth SDSS Data Release cover the entire OSIRIS field, with the exception of the central area corresponding to the location of the most massive stars. The SDSS catalog is therefore not useful to study the central cluster, and it is also 3 magnitudes shallower than the OSIRIS catalog, but it does provides useful information such as the magnitudes in $u$ and $g$ bands important for studying stellar photospheric properties, and important means for verifying the photometric calibration. \par
	We matched the OSIRIS and SDSS DR8 data in two steps. The first step was aimed to define the correct matching radius, by experimenting with increasing radii from $0.1^{\prime \prime}$ to $2^{\prime \prime}$ and comparing the {\it differential} distributions of observed, real, and spurious matches at increasing matching radii. The number of spurious coincidences has been calculated as in \citet{Dami06}, taking into account that the two catalogs are correlated; the total of real matches is given by the difference between the observed and spurious coincidences.  We adopted a final matching radius of $0.3^{\prime \prime}$, which is the highest value at which the differential distribution of spurious matches is negligible (few percent) compared with that of real matches. At this radius, we found 10790 matches between the OSIRIS and IPHAS catalogs, with 54 expected spurious matches and 8 multiple matches where a source of one catalog is matched with more than one of the other. After correcting for residual zero-points and color dependence in each OSIRIS field, we achieved negligible median offsets between the OSIRIS and SDSS magnitudes smaller than $0.003^m$. \par

	\subsection{Cross-correlation with the IPHAS catalog}
	\label{ipahs_sec}

	Another catalog in three optical bands ($r^{\prime}$, $i^{\prime}$, and H$\alpha$) of the Cyg~OB2 region can be extracted from the observations taken with the Wield Field Camera (WFC) on the 2.5m Isaac Newton Telescope (INT) for the INT Photometric H$\alpha$ Survey (IPHAS, \citealt{Dre05}). A total of 11995 IPHAS sources fall in the OSIRIS field. Our new OSIRIS data are about 5 magnitudes deeper in $r$ and $i$ than the IPHAS catalog. However, the latter is 4 magnitudes brighter than our saturation limit in the $r$ band. A combination of both catalogs allows us the study of stars over a wide range of magnitude, from $r=12^m$ to $r=25^m$. One complication is that, as described in Sect. \ref{cal_sect}, our OSIRIS catalog has been calibrated in the SDSS photometric system (AB), while the IPHAS catalog is calibrated in the Vega system. Suitable color transformations between the two catalogs have to be found in order to convert the IPHAS colors into the AB system. We adopted an empirical approach whereby the color transformation was obtained from sources common to both catalogs. \par
To match the IPHAS data with the SDSS+OSIRIS catalog, we adopted the same procedure described in the previous section, using a matching radius equal to $0.5^{\prime \prime}$, and finding 11199 matches between the OSIRIS and IPHAS catalogs (17 expected spurious matches), divided in 6104 sources with OSIRIS, IPHAS and SDSS counterparts and 5095 with only OSIRIS and IPHAS counterparts. \par
	A set of photometric transformations between the SDSS and WFC photometric systems has been already found by \citet{Gonzalez2011}:

\begin{equation}
r^{\prime}_{WFC}=r_{SDSS}-0.144+0.006\times \left( g-r \right)_{SDSS}
\end{equation}
\begin{equation}
i^{\prime}_{WFC}-i_{SDSS}-0.411-0.073\times \left( r-i \right)_{SDSS} 
\label{transf_eqGS}
\end{equation}

	 with a weak color term in $r$ band given by the similarity of the two used $r$ filters. Since the SDSS+IPHAS data of the Cygnus field cover a larger range of colors ($g-r<4^m$) than those used by \citet{Gonzalez2011} ($g-r<1.8^m$), we reviewed these transformations, selecting stars with errors in the relevant colors smaller than $0.15^m$ and in the magnitude offsets smaller than $0.03^m$. We also removed in four iterations the sources with an offset more deviant than $3\sigma_{offset}$ from the value expected from the linear interpolation of the data. The resulting transformations are slightly different than those found by \citet{Gonzalez2011}: 

\begin{equation}
r^{\prime}_{WFC}=r_{SDSS}-0.105-0.034\times \left( g-r \right)_{SDSS}
\label{transf_eq2a}
\end{equation}
\begin{equation}
i^{\prime}_{WFC}-i_{SDSS}-0.351-0.071\times \left( r-i \right)_{SDSS} 
\label{transf_eq2b}
\end{equation}
	
	mostly for the presence of a significant color term in the $r$ band, which is very likely induced by the small difference in extinction at the central wavelengths of the two $r$ filters (see, for instance, \citet{Donne94}), whose effect is dominant only for very red objects. \par
\begin{figure}[]
	\centering	
	\includegraphics[width=8.5cm]{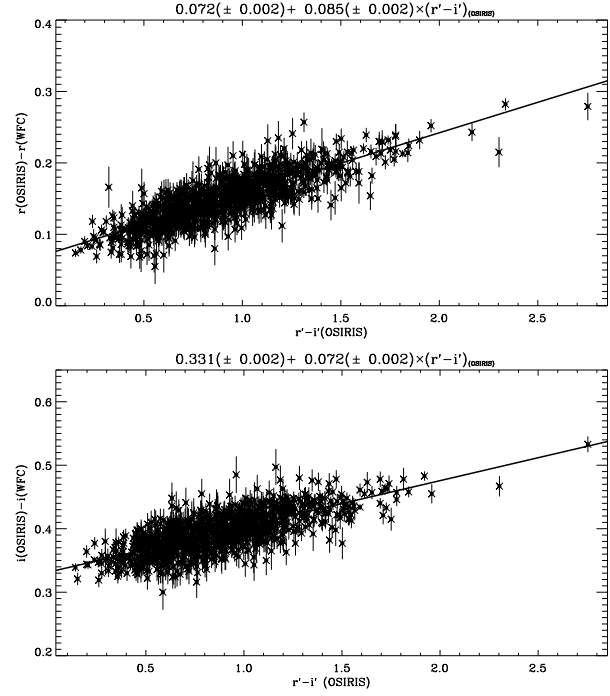}
	\caption{Magnitude differences in $r_{OSIRIS}-r^{\prime}_{IPHAS}$ and $i_{OSIRIS}-i^{\prime}_{IPHAS}$ versus the OSIRIS $r-i$ color for the stars in common between the OSIRIS and IPHAS catalogs selected as described in the text. The continuous lines, whose coefficients are shown in the top of each plot, are obtained from a linear fit to the data. The dotted line in the top panel marks the median $r$ offset.}
	\label{matchmag_image}
	\end{figure}

To convert the WFC magnitudes into the SDSS system we used the $r-i$ color in both magnitudes, expecting to find a significant color dependence like those in Eq. \ref{transf_eq2a} and \ref{transf_eq2b}. Fig. \ref{matchmag_image} shows the magnitude offsets of $r_{OSIRIS}-r^{\prime}_{WFC}$ and $i_{OSIRIS}-i^{\prime}_{WFC}$ as a function of the OSIRIS $r-i$ color. The stars used in this diagram are the $good-photometry$ sources corresponding to single matches between the two catalogs, with $r$ and $r^{\prime}$ magnitude in the $14^m-18^m$ range, and with the error of the difference between the OSIRIS and IPHAS magnitudes smaller than 0.03$^m$ (which is the median difference error for all the matched sources). We also rejected in four iterations the stars whose offsets are more deviant than $3\sigma_{difference}$ from the value expected from the linear fit to the data (the fit was performed for each iteration). The coefficients obtained from the linear interpolation in the last iteration are shown in the top of each panel in Fig.~\ref{matchmag_image}. The transformations between the OSIRIS photometric system (the AB system) and the IPHAS one (the Vega system) that we obtained are then:

\begin{equation}
r_{OSIRIS}-r^{\prime}_{WFC}=0.072+0.085\times \left( r-i \right)_{OSIRIS}
\end{equation}
\begin{equation}
i_{OSIRIS}-i^{\prime}_{WFC}=0.331+0.072\times \left( r-i \right)_{OSIRIS} 
\label{transf_eq}
\end{equation}

where $r_{OSIRIS}$ and $i_{OSIRIS}$ are the OSIRIS and $r^{\prime}_{WFC}$ and $i^{\prime}_{WFC}$ the IPHAS magnitudes. Eq.~\ref{transf_eq} gives the following color transformation:

\begin{equation}
\left( r^{\prime}-i^{\prime} \right)_{WFC}=0.260+0.987\times \left( r^{\prime}-i^{\prime} \right)_{OSIRIS}
\label{col_eq}
\end{equation}

To verify these transformations, we used them to convert the expected $r^{\prime}-i^{\prime}$ color of a G2V star.  We adopted as the color value in the Vega system \citep{Dre05}, $\left(r^{\prime}-i^{\prime}\right)_{Vega}=0.384^m$, and in the AB system the value found by \citet{Fuku00}, $\left(r^{\prime}-i^{\prime}\right)_{AB}=0.127^m$, equal to $\left(r-i\right)_{AB}=0.126^m$ by using the conversion suggested in the SDSS DR8 website\footnote{http://www.sdss.org/dr7/algorithms/jeg\_photometric\_eq\_dr1.html\#usno2SDSS}. Using Eq. \ref{col_eq} to convert the color of a G2V star from the AB system to the Vega system, we find 0.384$^m$, reproducing properly the Vega $r-i$ color of a solar type star.

	\subsection{Cross-correlation with the {\it Chandra} catalog}
	\label{chan_sec}

Thanks to the high level of X-ray emission characterizing young PMS stars \citep[e.g.][]{Mont96} X-ray selection provides a powerful means to isolate the cluster population from the myriad foreground and background stars. The combination of our optical catalog with X-ray data allows us to identify and study the cluster loci in the optical diagrams.  Two existing {\it Chandra} ACIS-I observations partially cover the field observed with OSIRIS. The first one (P.I.: Flaccomio) centered on Cyg~OB2 ($\alpha=20:33:11.0$ and $\delta=+41:15:10.00$) with a total exposure time of $99.7\,ks$ was first analyzed by \citet{Alba07}. The second one (P.I.: Butt) covers a field to the north-west, centered on $\alpha=20:32:07.0$ and $\delta=+41:30:30.00$, with an exposure time of $49.4\,ks$. Both observations have been analyzed by \citet{Wri09}, who derived a catalog of 1696 X-ray sources complete down to $1\,M_{\odot}$. The FoVs of the two {\it Chandra} observations are shown in Fig. \ref{fov_image}. \par
	We first performed a preliminary match between the optical and X-ray catalogs by using a $2^{\prime \prime}$ matching radius, in order to determine and correct systematic position offsets.   Since the PSF in the ACIS-I detector (and the resulting instrument sensitivity) degrades from the center of the  field outward, for the final match we used a variable matching radius.  Moreover, the procedure adopted in Sect. \ref{ipahs_sec} to derive the distribution of spurious matches at different matching radii requires the matched catalogs to be uniformly distributed in the common area. This is a good approximation for the optical catalogs, but, owing to the variable sensitivity of the {\it Chandra} FoV, the spatial distribution of {\it Chandra} sources can not be considered uniform. \par
To account for the varying PSF, each X-ray source has been matched with close optical sources by using a matching radius proportional to its positional uncertainty: $r_{match}=A\times \sigma_{pos}$. The proportional constant $A$ is common for every X-ray source, and it has been found by comparing the distributions of spurious and real matches obtained with 20 values of $A$ ranging from 0.1 to 5. For each value of $A$, we performed 4 matches, applying a rigid translation to the X-ray catalog in order to derive the number of expected spurious matches $N_{spurious}\left( A \right)$. The total number of matches $N_{total}\left( A \right)$ has been derived by matching the X-ray catalog without translations, and the real number of matches $N_{real}\left( A \right)=N_{total}\left( A \right)-N_{spurious}\left( A \right)$. Comparing again the distributions of real and spurious matches, we set the value of $A$ equal to 1.5, this being the largest value at which the differential distributions of spurious coincidences is negligible with respect to that of real matches. With this definition of the individual matching radius, we obtained a total of 1407 matches, with 22 expected spurious matches. This result is a great improvement over the catalog published by \citet{Wri09}, who matched the X-ray and existing IPHAS catalog of this region finding 750 coincident sources.

\section{Visual extinction in Cyg~OB2}
\label{av_sec}

	\subsection{Extinction in the direction of the association}
	\label{av1_sec}

Despite the proximity of Cyg~OB2 compared with other massive clusters, it is affected by a large visual extinction due both to the dust associated with the Great Cygnus Rift in the foreground and its parental cloud. The most recent estimates of the visual extinction in the direction of Cyg~OB2 are those of \citet{Sale09}, who studied how $A_V$ varies with distance in the direction of the association and found an increase of $A_V$ from $2^m$ to $5^m$ in the region between $1\,kpc$ and $2\,kpc$ away; \citet{Dre08}, who found that the cluster locus in the $r^{\prime}-H\alpha$ vs. $r^{\prime}-i^{\prime}$ exhibits an extinction between $A_V=2.5^m$ and $A_V=7^m$, with a larger extinction in the area surrounding the association; and \citet{Wri00}, who found a median $A_V=7.5^m$ in the central cluster and $A_V=5.5^m$ in the north-west field observed by {\it Chandra}. \par
\begin{figure}[!ht]
	\centering	
	\includegraphics[width=8.5cm]{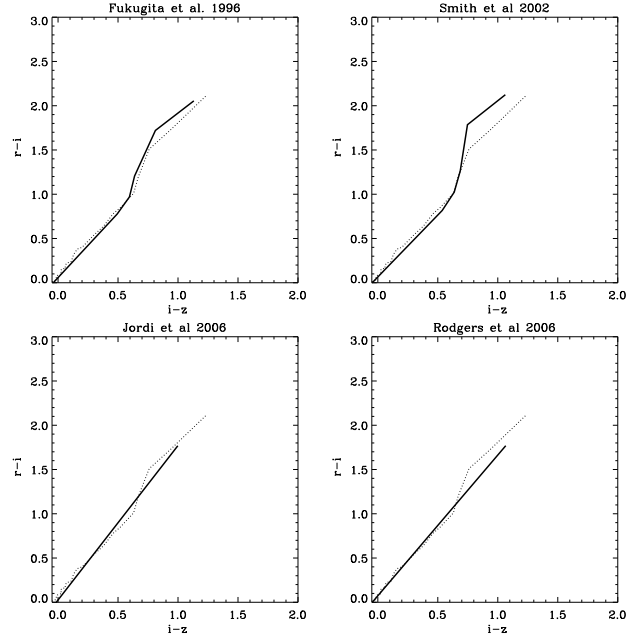}
	\caption{$r-i$ and $i-z$ colors of the Main Sequence stars with different spectral types observed \citet{Cove07} (dotted line), and those predicted by the ZAMS of \citet{Sie00}, with different transformation from the $UBVRI$ to the $ugriz$ photometric system applied.}
	\label{transformations}
	\end{figure}

	Since colors are not affected by distance, the color-color $r-i$ vs. $i-z$ diagram is suitable for estimating the visual extinction affecting Cyg~OB2 members, by fitting the isochrones from \citet{Sie00} to the cluster locus. However, the photometric data of the isochrones are in the Johnson-Cousin $UBVRI$ photometric system, so they must be converted using proper transformations from the $UBVRI$ to the $ugriz$ system. Several transformation have been published. In order to choose the most reliable for our case, we compared the $r-i$ and $i-z$ colors of the ZAMS from \citet{Sie00} converted using different transformations with those of the main sequence stars observed by \citet{Cove07}, as shown in Fig. \ref{transformations}. The transformation that reproduces better the colors of the Covey MS stars is that from \citet{Fuku96}. This transformation converts the $UBVRI$ photometry into the $u^{\prime}g^{\prime}r^{\prime}i^{\prime}z^{\prime}$ system, so it is necessary to apply also the transformation between the $u^{\prime}g^{\prime}r^{\prime}i^{\prime}z^{\prime}$ and the $ugriz$ systems. The transformation from \citet{Smi02} fails to reproduce the colors of late stars. The transformations from \citet{Jordi06} and \citet{Rodgers06} transform the Siess ZAMS into a straight line. The former is a better approximation of the irregular shape of the Main Sequence from \citet{Cove07}, but the difference between the two transformations is small. For these reasons, we mainly adopt the \citet{Fuku96} transformations. \par
\begin{figure}[]
	\centering	
	\includegraphics[width=7.5cm]{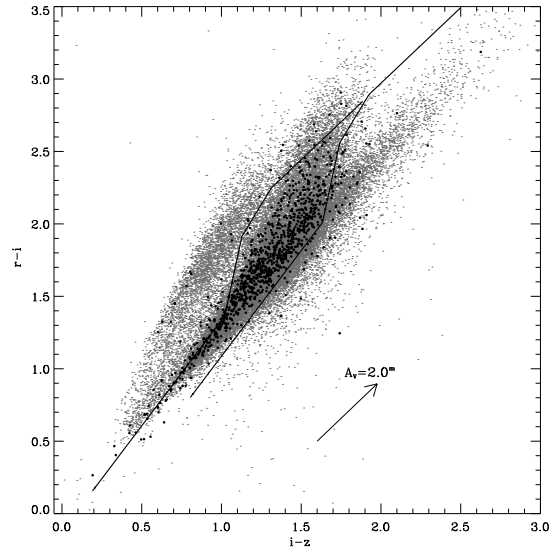}
	\caption{$r-i$ vs. $i-z$ diagram of the $good-photometry$ stars in the whole OSIRIS FoV (gray points). The black dots mark the colors of optical sources with X-ray counterparts.  The black lines are the $3.5\,Myr$ isochrone from \citet{Sie00}, drawn for two different values of extinction ($A_V=2^m$ and $A_V=6^m$).}
	\label{colcol_image}
	\end{figure}

Fig.~\ref{colcol_image} shows the $r-i$ vs. $i-z$ diagram of all the $good-photometry$ stars in the OSIRIS FoV. Since almost all the optical sources with X-ray counterparts are young cluster members, they define a well determined locus in the diagram.  In Fig.~\ref{colcol_image} we also show the colors predicted by the $3.5\,Myr$ isochrone \citep{Sie00}, which is the mean age of the stars in the center of Cyg~OB2 \citep{Wri00}, transformed into the $ugriz$ system and reddened by using the extinction law derived by \citet{Donne94}. From the comparison of the observed colors with this isochrone we find the extinction of the cluster members to range mainly from $A_V=2^m$ to $A_V=6^m$. A few X-ray sources show higher extinctions, being largely embedded cluster members or background sources. A large group of sources lies to the left of the $3.5\,Myr$ isochrone with $A_V=2^m$, and is separated from the cluster locus by a small gap with a smaller density of stars compared with the rest of the diagram, being mostly in the foreground. \par
\begin{figure}[!ht]
	\centering	
	\includegraphics[width=6cm]{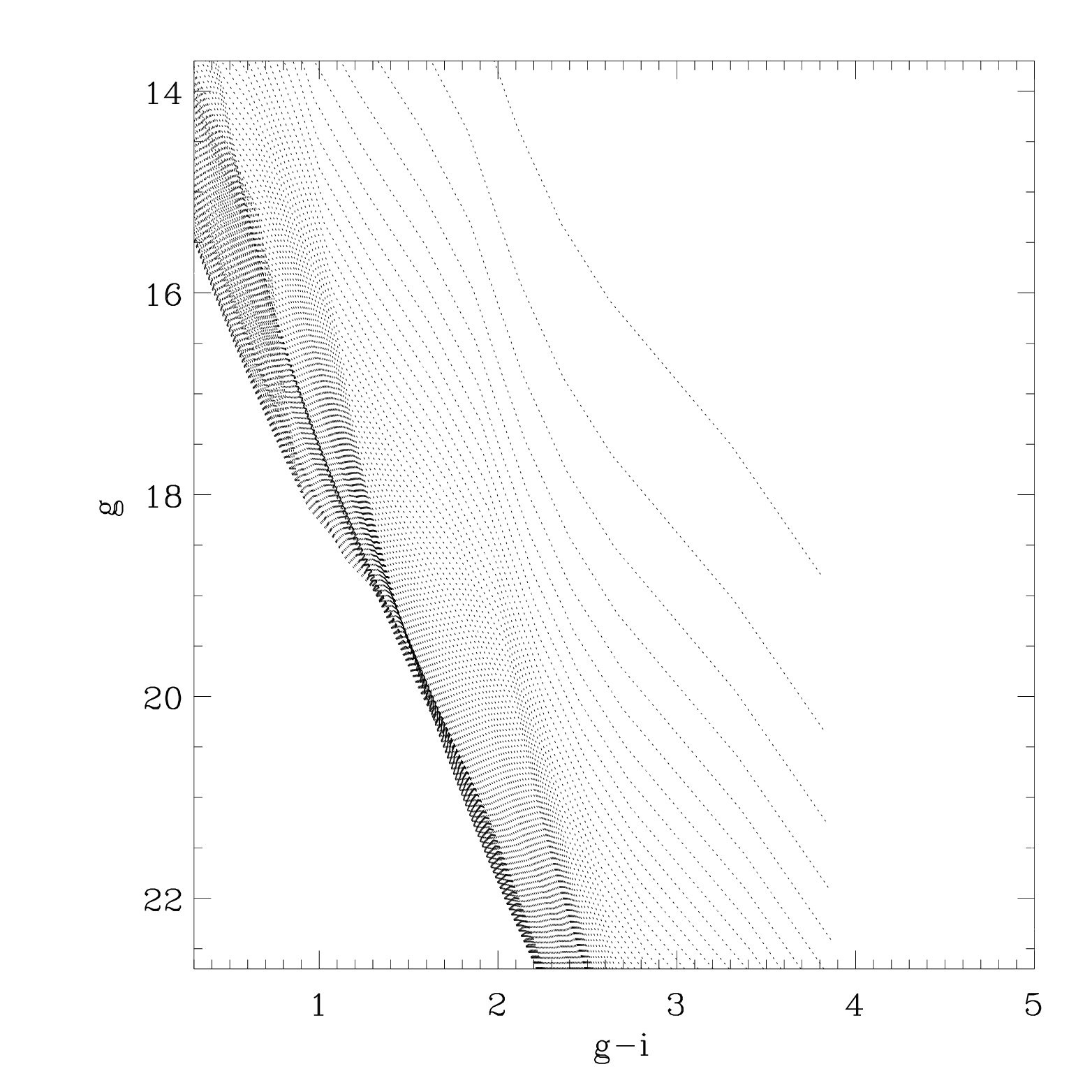}
	\includegraphics[width=6cm]{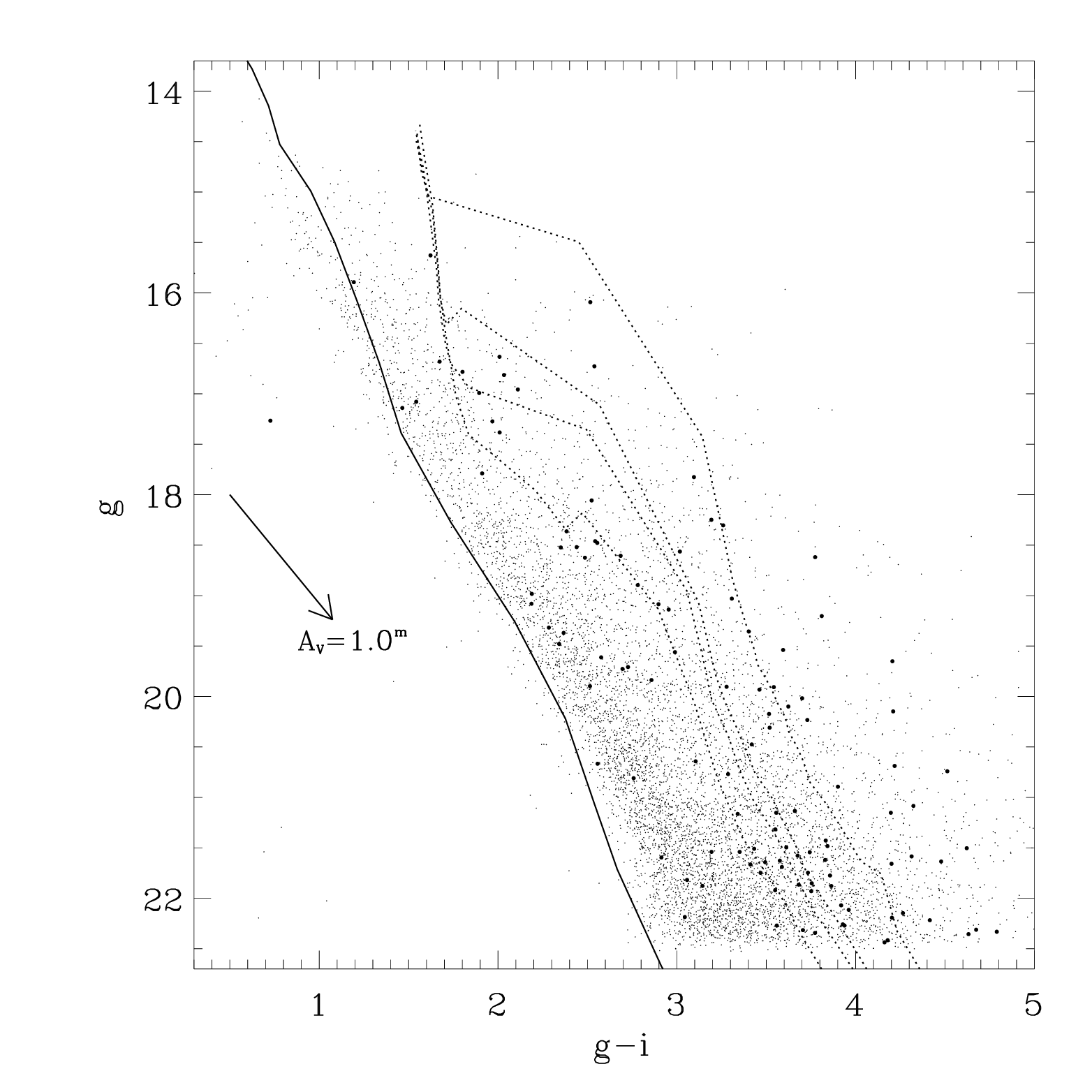}
	\caption{$g$ vs. $g-i$ diagrams of 100 ZAMS from \citet{Sie00}, drawn adopting increasing distance and visual extinction (left panel); and of the SDSS-OSIRIS sources falling in the OSIRIS field (small dots), stars with X-ray counterpart (large dots), the ZAMS from \citet{Sie00} with $A_V=1^m$ and a distance of $850\,pc$ (black line), and the isochrones with age ranging from $0.5\,Myrs$ to $10\,Myrs$ with distance and median extinction typical of cluster members (right panel).}
	\label{ggi}
	\end{figure}

	In principle, the cloud responsible for the extinction affecting Cyg~OB2 can be in the foreground or associated with the cluster. However, there are several indications that Cyg~OB2 is not yet embedded in the parental cloud \citep{Sch06} and it lies behind the Cygnus Rift, which is responsible for a steep rise of the visual extinction along this direction. Since the optical colors at short wavelength are largely affected by extinction, it is possible to estimate the distance of the nebulosity which is mainly responsible for the rise of visual extinction from the color-magnitude diagram $g$ vs. $g-i$. The left panel of Fig. \ref{ggi} shows 100 ZAMS from \citet{Sie00} in this color-magnitude diagram, each drawn adopting a visual extinction $A_V=0.5 \times dist(kpc)$, typical of lines of sight crossing an uniform and loose distribution of interstellar medium, and distance ranging from $500\,pc$ to $5500\,pc$. As expected, increasing distance and extinction the ZAMS populate the lower part of the diagram. The right panel of Fig. \ref{ggi} shows the observed $g$ vs. $g-i$ diagram of the OSIRIS-SDSS sources with good photometry (small dots), with those detected in X-rays marked by large dots. The dashed lines are isochrones from \citet{Sie00} drawn with the distance adopted for Cyg~OB2 ($1450\,pc$) and the average extinction of $A_V=4.2^m$ (see Sect. \ref{av2_sec}). Comparing directly the two panels, it is evident that the observed diagram is less populated by faint blue sources than what expected from the left panel. This is due to the presence of the nebulosity along the line of sight, which causes a steep rise of the reddening of the stars behind it. The distance of this cloud can be estimated, then, by finding what are the values of $A_V$ and distance at which the ZAMS fits the blue end of the observed distribution of stars in the $g$ vs. $g-i$ diagram. The value of distance at which we obtained the best fit is $850\pm25\,pc$, with $A_V=1^m$. This value is only slightly smaller than the distance found by \citet{Sale09}, who estimate that the extinction along this line of sight starts to increase from $A_V=2^m$ ($\sim 1\,kpc$) to $A_V=5^m$ ($\sim 2\,kpc$). Our estimate also confirms that the nebulosity responsible for the high visual extinction is not associated with Cyg~OB2, but is in the foreground.

	\subsection{Individual extinction of candidate cluster members}
	\label{av2_sec} 

Taking advantage of the fact that the $3.5\,Myr$ isochrone in the $r-i$ vs. $i-z$ diagram can be transformed into a straight line with the transformations of \citet{Jordi06} with good approximation, it is possible to calculate the individual extinction of the candidate cluster members from the the displacement of the X-ray sources from the $3.5\,Myr$ isochrone drawn with $A_V=0^m$ along the reddening vector.  Indeed, we verified that contributions to these displacements from the expected range of ages of Cyg~OB2 cluster members are negligible. In this calculation, we excluded the few sources more red than the colors of a $3.5\,Myr$ old B2.5 star at different extinctions.  The variation in apparent extinction could be caused by a patchy distribution of gas and dust along the line of sight, or by intrinsic circumstellar gas such as a protoplanetary disk. The presence of a circumstellar disk is only expected to significantly affect the {\em optical} colors of stars when intense accretion is ongoing, or for a disk inclination larger than $80^{\circ}$ with respect to the line of sight \citep{Io10}.  For most sources then it is likely that the variations in extinction correspond to an inhomogeneous distribution of gas and dust in the cluster environment. \par
\begin{figure}[]
	\centering	
	\includegraphics[width=7.5cm]{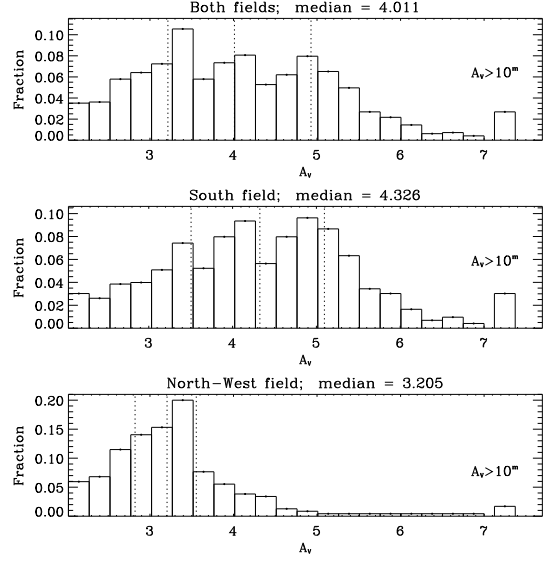}
	\caption{Histograms showing the distribution of total extinction for optical-X-ray sources in both {\it Chandra} fields and in each field alone. The vertical dotted lines mark the median value and the 25\% and 75\% quartiles. }
	\label{avisto_image}
	\end{figure}

	Fig.~\ref{avisto_image} shows the distributions of individual extinctions calculated for the optical sources with X-ray counterparts falling in both {\it Chandra} fields (upper panel) and in each field (lower panels). We have not considered sources with $A_V<2^m$, since they are more likely foreground stars (a total of 56 objects). Sources in the southern {\it Chandra} field (centered on Cyg~OB2) have a median extinction equal to $4.33^m$. This median extinction is smaller than the value found by \citet{Wri00} with an isochrone-fitting technique, but it is  within the range of extinction between $1\,kpc$ and $2\,kpc$ found by \citet{Sale09} ($2<A_V<5$) and \citet{Dre08} ($2.5<A_V<7$). In the histogram (middle panel) the large differential reddening affecting the central cluster is also evident, since the distribution does not show an evident peak and the bins from $A_V=3.0^m$ to $A_V=5.0^m$ are almost equally populated. The extinction distribution in the north-west field is quite different, with a median value equal to $3.21^m$, and a well defined peak with no long tail toward larger extinction.  This indicates that the density of the obscuring material changes significantly in the direction of the two fields. \par
\begin{figure*}[]
	\centering	
	\includegraphics[width=6.5cm]{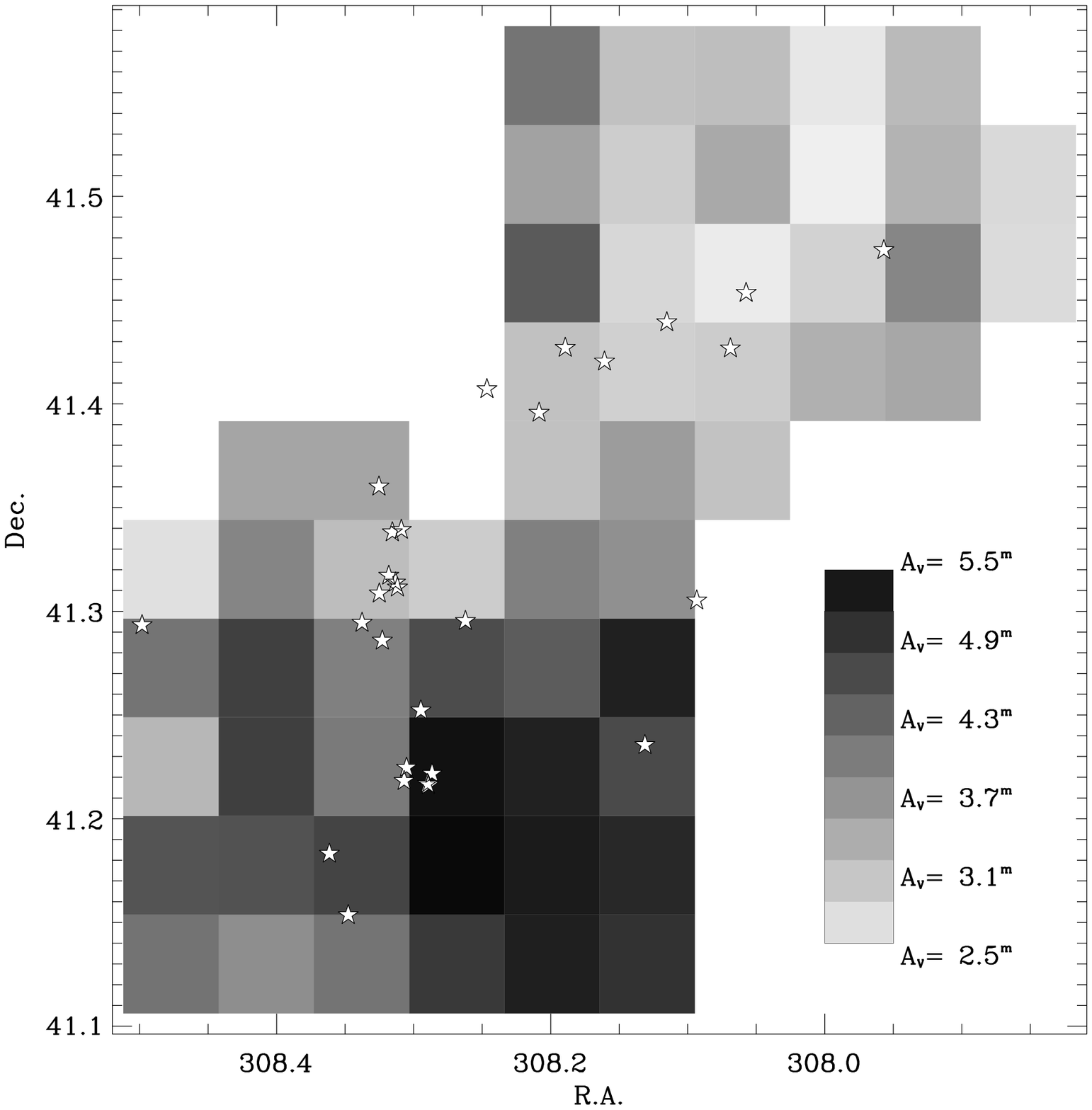}
	\includegraphics[width=6.5cm]{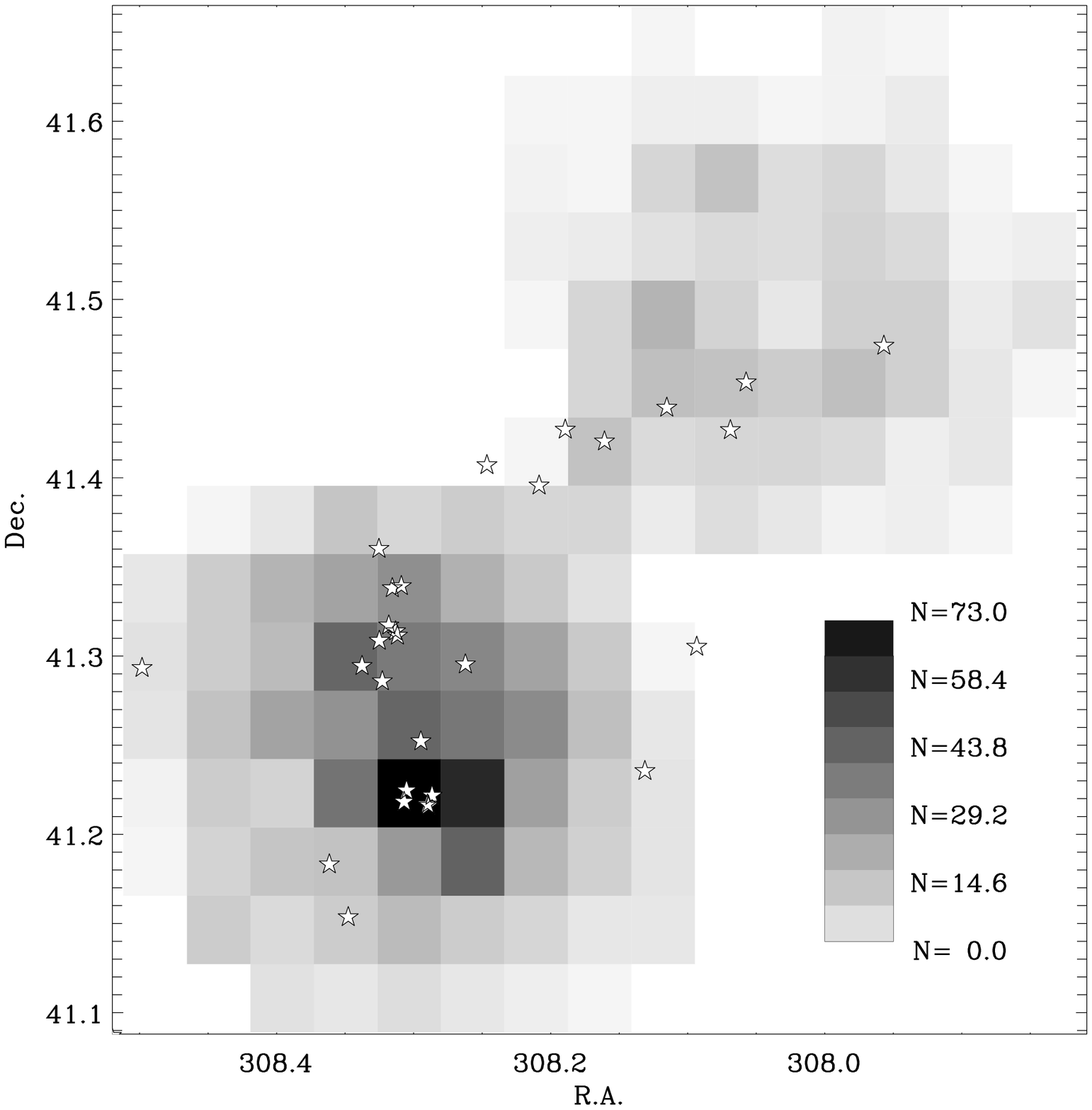}
	\caption{Left panel: gray-scale map of the median extinction for the X-ray sources with optical counterparts with $A_V>2^m$ falling in each spatial bin.  Right panel: density map of the candidate cluster members.  Stars mark the positions of the cluster members with O spectral type.}
	\label{avmap_image}
	\end{figure*}

To better visualize how the extinctions vary across the field, the left panel of Fig. \ref{avmap_image} shows a gray-scale map of the visual extinction, where the $A_V$ value characterizing each bin is the median value of the extinction of the sources falling in the bin.  The median number of sources in each bin is 11, with the less populated bins lying along the border of each ACIS field.  The locations of the known O stars are also indicated.  In the north-west field the extinction distribution is almost uniform, while in the southern field the extinction increases in the center-west direction. \par
	The right panel of Fig.~\ref{avmap_image} shows the density map of the candidate cluster members (optical+X-ray sources with $A_V>2^m$). In the southern field the density of cluster members, both low-mass and O stars, clearly peaks in the center of the field, corresponding to one of the most extinguished regions. This could hint that, despite the large content of massive stars, Cyg~OB2 is still embedded in the parental cloud.  However, the CO map shown in \citet{Sch06} suggests that the cluster has created a cavity in the parental cloud,  as is commonly observed in massive clusters in our Galaxy with an age of a few Myrs.  On the other hand, in this CO map small regions of higher density lie close to the positions of the two groups of massive stars and could be responsible for the peak of extinction evident in Fig.~\ref{avmap_image}. The selection of the members with disks, with the identification of those members still embedded in dense regions of the parental cloud, will shed some light on the possibility that embedded sources and dense intracluster medium is still present in the center of Cyg~OB2. \par

\section{Color-magnitude diagrams of Cyg~OB2 members}
\label{cm_sec}

\begin{figure*}[]
	\centering	
	\includegraphics[width=16cm]{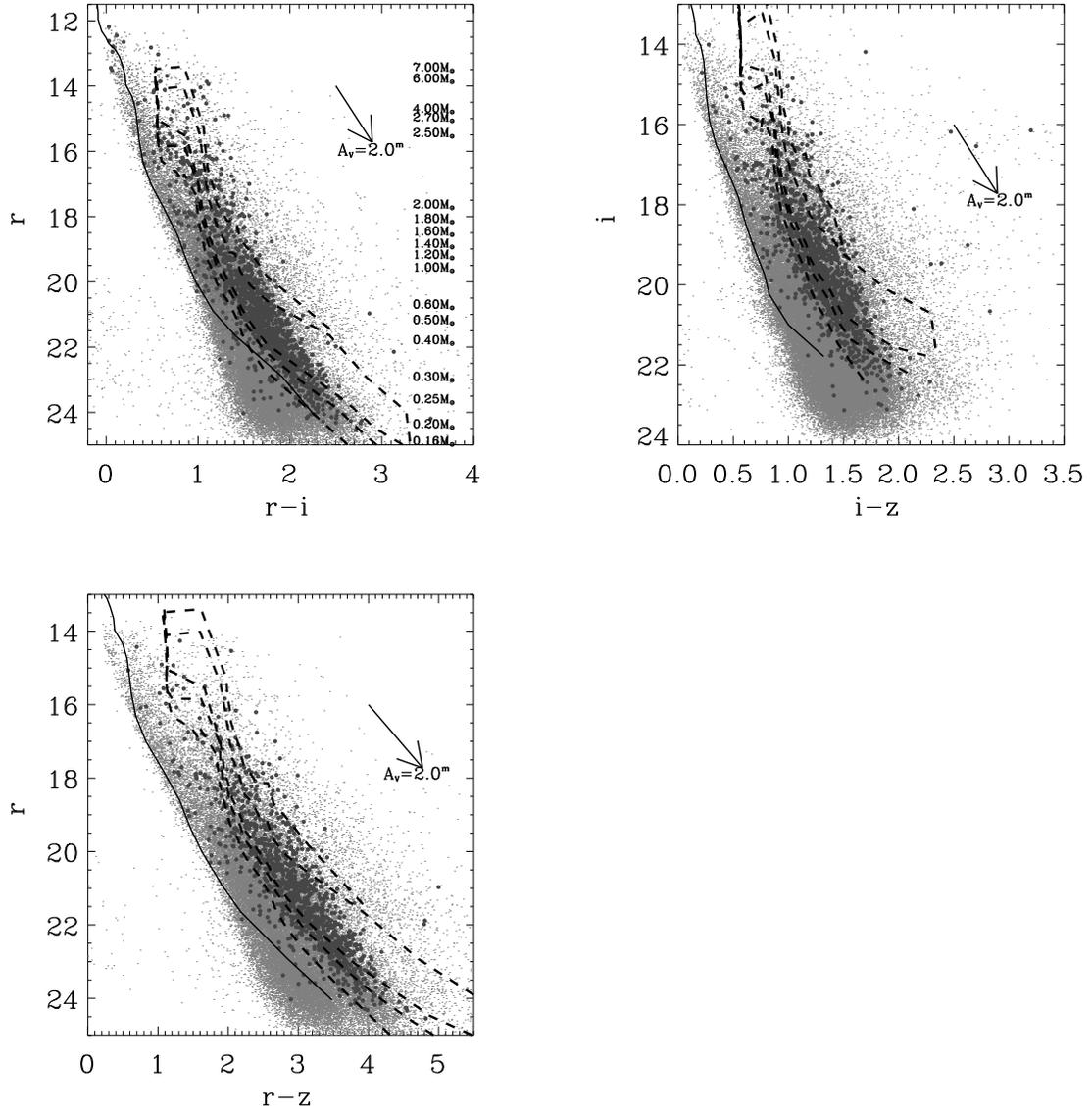}
	\caption{Color-magnitude diagrams of the $good-photometry$ stars in the OSIRIS FOV (small gray dots). The black line is the ZAMS at a distance of $850\,pc$ and an extinction $A_V=1^m$. The dashed lines are the isochrones for stars of 0.5, 1, 3.5, 5, and 10$\,Myr$, at the distance but with an extinction $A_V=4^m$. The large darker dots are the optical sources with X-ray counterparts. The masses listed on the right side of the diagram in the left panel are those predicted by a $3.5\,Myr$ isochrone.}
	\label{diag_image}
	\end{figure*}

Fig.~\ref{diag_image} shows the three color-magnitude diagrams of the $good-photometry$ sources in the whole OSIRIS field.  Stars with X-ray counterparts are highlighted.  In all the diagrams, the X-ray sources define a well delimited cluster locus.  By drawing the isochrones younger than $10\,Myr$ with an extinction of $A_V=4.2^m$, the cluster loci in all the color-magnitude diagrams mostly lie between the $5\,Myr$ and $1\,Myr$ isochrones. The ZAMS, drawn adopting the distance of the foreground cloud of $850\,pc$ and an extinction of $A_V=1^m$, is well fitted to the blue end of the locus of main sequence foreground stars, indicating that these parameters are appropriate. Most of the X-ray sources bluer than the $10\,Myr$ isochrone are those with extinction smaller than $2^m$ that are very likely in the foreground. \par
	 The $r$ vs. $r-i$ diagram in Fig.~\ref{diag_image} is completed using the IPHAS data, converted into the SDSS photometric system by using the transformation we found in Sect.~\ref{ipahs_sec}; all the three diagrams are also completed with the SDSS data. Note that, thanks to the depth of the OSIRIS observations, we now have optical photometry for a significant number of faint low-mass candidate cluster members: 484 X-ray sources with $r>21.5^m$ out of a total number of 1250 sources with both X-ray and optical counterparts with good photometry. This will be very important for future studies based on the selection of low-mass cluster members and the effects of massive stars on their early evolution.  The masses listed on the right of the diagram, corresponding to the $3.5\,Myr$ isochrone (representative of the median age of the central cluster as noted earlier; \citealp{Wri00}) demonstrate that the grasp of the OSIRIS photometry is deep enough to observe stars with sub-solar mass at the distance of Cyg~OB2.\par
Among the IPHAS sources with X-ray counterparts, those with $r-i\sim0.5^m$ and $12^m<r<16^m$ are compatible with B stars with only moderate extinction. Most of them are in the north-west field, where the extinction is lower in general, as might be expected. A group of X-ray sources with $r-i\sim0.1^m$ and $13.5^m<r<12^m$ are compatible with being foreground objects.  This is also consistent with their positions  in the $r-i$ vs. $i-z$ and $r^{\prime}-H\alpha$ vs. $r^{\prime}-i^{\prime}$ diagrams and with their low neutral hydrogen column densities, $N_H$, deduced from their X-ray spectra  that are compatible with extinctions $A_V<0.2^m$ \citep{Wri00}.  Surprisingly, their spatial distribution is not sparse as expected for uncorrelated foreground objects, but instead they all lie in the north-west field. \par

\section{Conclusions}

We have analyzed new GTC/OSIRIS optical observations in $r^{\prime}i^{\prime}z^{\prime}$ bands of a field of size $41^{\prime}\times 41^{\prime}$ approximately centered on Cyg~OB2. The resulting catalog contains the photometry of 64157 optical sources, among which 38300 have good photometry in all three bands. The catalog reaches $r^{\prime}=25^m$, 5 magnitudes deeper than the existing IPHAS catalog and 3 than the SDSS DR8 data. This limit corresponds to a $3.5\,Myr$ star with $M=0.15\,M_{\odot}$ at the adopted distance and extinction of Cyg~OB2. The stellar density of stars with good photometry varies significantly from north southward, with the maximum density corresponding to a cavity in the cloud in the north-west part of the observed area. \par
	We cross-correlated the OSIRIS catalog with an existing {\it Chandra} ACIS-I catalog, of the central and north-west area of the field, in order to define the cluster locus in the diagrams. This yielded 1407 optical sources with X-ray counterparts, almost twice the number of X-ray sources with optical counterparts identified in previous works in these fields. Using the stars with detections in both the OSIRIS and IPHAS catalogs, we derived a suitable color transformation between the IPHAS and OSIRIS photometric systems (Vega and AB, respectively). This transformation converts properly the colors of a G2V star from the AB to the Vega system. \par
The $r-i$ vs. $i-z$ color-color diagram shows a clear cluster sequence which can be encompassed by a $3.5\,Myr$ isochrone with extinction between $2^m$ and $6^m$.  This extinction turns out to be the main range of extinction of cluster members.  We also found that the mean extinction decreases from the central cluster northward: in the central field we found a median extinction $A_V=4.33^m$, with a large differential reddening across the field; while in the north-west {\it Chandra} field we found a median extinction $A_V=3.21^m$, with a more peaked extinction distribution.  We derived an extinction map of the region that reveals evidence that the most obscured regions are also those with the largest density of low-mass and massive member stars. \par
	The color-color diagram also exhibits a gap between the foreground objects and the cluster locus that is induced by a steep increase of the visual extinction. This is due to the presence of a dense nebulosity along this line of sight. The fit of the zero age main sequence with the foreground population in the $g$ vs. $g-i$ diagram suggests a distance of $850\pm25\,pc$ for this cloud, which turns out to be in the foreground and not associated with Cyg~OB2.


\acknowledgments
This article is based on observations made with the Gran Telescopio CANARIAS (GTC), installed in the Spanish Observatorio del Roque de los Muchachos of the Instituto de Astrof\'{i}sica de CANARIAS, in the island of La Palma, and it also makes use of data obtained as part of the INT Photometric Halpha Survey of the Northern Galactic Plane (IPHAS) carried out at the Isaac Newton Telescope (INT) (all IPHAS data are processed by the Cambridge Astronomical Survey Unit, at the Institute of Astronomy in Cambridge) and from archival Chandra/ACIS-I observations. MGG and NJW were supported by Chandra Grant GO0-11040X. JJD, VLK, and TA were supported by NASA contract NAS8-39073 to the {\it Chandra} X-ray Center (CXC) and thank the Director, Harvey Tananbaum, and the CXC science staff for advice and support. DGA acknowledge support from the Spanish MICINN through grant AYA2008-02038. 


\newpage
\addcontentsline{toc}{section}{\bf Bibliografia}
\bibliographystyle{aa}
\bibliography{biblio}

\newpage

  	\begin{table}[]
	\centering
	\caption {Log of the observations}
	\vspace{0.5cm}
	\begin{tabular}{ccccc}
	\hline
	\hline
	Night& Fields & Standard Fields & Configuration & Seeing ($r^{\prime}$,$i^{\prime}$,$z^{\prime}$) ($arcsec$) \\
	\hline
	8/11/2009 &11-15	&PG1545; SP0346	&200/LN		&0.90, 0.94, 1.00		\\
	8/18/2009 &6-8	 	&GD71		&500/HN		&1.29, 1.21, 1.28		\\   
	8/20/2009 &1-5		&PG0231; SA92	&500/LN		&1.29, 1.21, 1.28		\\   
	8/21/2009 &8-10		&PG2317		&500/HN		&1.18, 1.12, 1.15		\\  
	8/21/2009 &18-22 	&PG2317		&500/LN		&1.08, 0.99, 1.04		\\   
	8/22/2009 &23-27 	&SA95		&500/LN		&1.47, 1.43, 1.48		\\  
	\hline
	\hline
	\multicolumn{5}{l}{} 
	\end{tabular}
	\label{log_tbl}
	\end{table}

  	\begin{table}[]
	\centering
	\caption {Coefficients of the photometric calibration}
	\vspace{0.5cm}
	\begin{tabular}{ccc}
	\hline
	\hline
	$r$& $i$ & $z$ \\
	\hline
	 &Left CCD zero-points& \\
	\hline
	3.58-4.00& 3.18-3.61 & 2.76-3.22 \\
	\hline
	\hline
	&Right CCD zero-points&\\
	\hline
	3.55-4.02& 3.15-3.61 & 2.80-3.29 \\
	\hline
	\hline
	&Color terms& \\
	\hline
	0.266-0.318& 0.065-0.087 & 0.013-0.029 \\
	\hline
	\hline
	\end{tabular}
	\label{coeff_table}
	\end{table}

\begin{table*}
\begin{minipage}[t][180mm]{\textwidth}
\caption{Part of the electronic catalog.}\label{cat_tab}
\centering
\begin{tabular}{ccccccccccc} 
\hline\hline  
ID&RA(J2000)&DEC(J2000)&$r^{\prime}$&$\sigma_{r^{\prime}}$&$i^{\prime}$&$\sigma_{i^{\prime}}$&$z^{\prime}$&$\sigma_{z^{\prime}}$&$\chi^2$&{\tt SHARP}\\
\hline
\hline 
      171&  308.6602478&  41.4834251&	    &	     & 24.193&        &  22.012&        &  0.00&  0.00\\
      172&  308.6459961&  41.4840431& 22.673&	0.012& 20.673&  0.005 &  18.940&   0.006&  1.26&  0.50\\
      173&  308.6399231&  41.4851074& 24.659&	0.159& 22.822&  0.037 &  21.072&   0.028&  1.19&  1.13\\
      174&  308.6605835&  41.4853439& 25.513&	0.327& 23.922&  0.103 &  21.748&   0.126&  1.25&  4.56\\
      175&  308.6415100&  41.4851723& 22.106&	0.010& 20.688&  0.007 &  19.425&   0.014&  1.16&  0.37\\
      176&  308.6547546&  41.4855270&	    &	     & 23.922&        &  22.405&        &  0.00&  0.00\\
      177&  308.6530151&  41.4855690& 21.426&	0.007& 19.808&  0.005 &  18.920&   0.006&  1.20&  0.15\\
      178&  308.6368103&  41.4860077& 24.532&	0.122& 22.391&  0.021 &  19.908&   0.009&  1.29&  1.32\\
      179&  308.6352539&  41.4861221& 24.496&	0.109& 22.598&  0.039 &  21.137&   0.040&  1.21&  1.47\\
      180&  308.6576538&  41.4866714& 21.586&	0.007& 20.192&  0.005 &  18.933&   0.011&  1.17&  0.10\\
      181&  308.6414185&  41.4865875& 23.498&	0.036& 21.873&   0.015&  20.413&   0.013&  1.19&  0.18\\
      182&  308.6518250&  41.4867249& 24.573&	0.087& 22.790&   0.057&  21.194&   0.036&  1.18&  0.46\\
      183&  308.6606445&  41.4868431& 25.838&	0.391& 23.777&   0.108&  21.720&   0.036&  1.21&  1.36\\
      184&  308.6435242&  41.4867668& 22.842&	0.017& 21.262&   0.009&  19.937&   0.014&  1.17&  0.42\\
      185&  308.6415710&  41.4869843& 24.671&	0.070& 22.630&   0.038&  21.058&   0.019&  1.15&  0.39\\
      186&  308.6609802&  41.4873161& 25.137&	0.289& 24.342&   0.213&  22.209&   0.168&  1.21&  1.32\\
      187&  308.6427612&  41.4873199& 23.695&	0.033& 21.560&   0.015&  19.805&   0.011&  1.19&  0.18\\
      188&  308.6546021&  41.4875298& 18.918&	0.005& 17.131&   0.001&  15.452&   0.001&  1.62& -0.02\\
      189&  308.6382141&  41.4877434&	    &	     & 24.594&        &  22.874&        &  0.00&  0.00\\
      190&  308.6494446&  41.4878654& 24.681&	0.189& 24.275&   0.215&  21.945&   0.054&  1.18&  2.64\\

\hline
\multicolumn{11}{l}{If both $\chi^2$ and {\tt SHARP} are equal to 0 then the source was observed in only one optical band,} \\ 
\multicolumn{11}{l}{and it was not possible to calibrate its magnitude.}
\end{tabular}
\vfill
\end{minipage}
\end{table*}

\newpage
\par

\end{document}